\documentclass[a4paper]{jpconf}
\usepackage{graphicx} 
\begin{document}
\title{Quantum Foam, Gravitational Thermodynamics, and the Dark Sector}

\author{Y. Jack Ng}

\address {Institute of Field Physics, Department of Physics \& Astronomy,
University of North Carolina, Chapel Hill, NC 27599-3255, USA}

\ead{yjng@physics.unc.edu}

\begin{abstract}

Is it possible that the dark sector (dark energy in the form of 
an effective dynamical
cosmological constant, and dark matter) has its origin in quantum gravity?
This talk sketches a positive response. Here specifically quantum gravity 
refers to the combined effect 
of quantum foam (or spacetime foam due to quantum fluctuations
of spacetime) and gravitational thermodynamics.  We use two simple independent
gedankan experiments to show 
that the holographic principle
can be understood intuitively as having its origin in the quantum fluctuations 
of spacetime.  Applied to
cosmology, this consideration leads to a dynamical cosmological constant 
of the observed magnitude, a result that can also be obtained 
for the present and recent cosmic eras by using
unimodular gravity and causal set theory.  Next we  
generalize the concept of gravitational thermodynamics to a spacetime
with positive cosmological constant (like ours)
to reveal the natural emergence, in galactic dynamics, of a critical 
acceleration parameter related to the cosmological constant.  We are then led 
to construct a phenomenological model of dark matter which we call ``modified 
dark matter" (MDM) in which the dark matter density profile depends on
both the cosmological constant and ordinary matter. 
We provide observational tests of MDM 
by fitting the rotation curves to a sample 
of 30 local spiral galaxies
with a single free parameter and by showing that the dynamical and observed 
masses agree in a sample of
93 galactic clusters.  We also give a brief discussion 
of the possibility that quanta of both
dark energy and dark matter are non-local, obeying quantum
Boltzmann statistics (also called infinite statistics) as described by a 
curious average of the bosonic and 
fermionic algebras.  If such a scenario is correct, we can expect some
novel particle phenomenology involving dark matter interactions.  This may
explain why so far no dark matter detection experiments 
have been able to claim 
convincingly to have detected dark matter.

\end{abstract}

\section{Introduction}

This talk is  based on several loosely related pieces of work I did 
mostly with various
collaborators.  I will start with one aspect of John Wheeler's spacetime foam 
or quantum foam by
which he was referring to a foamy structure of spacetime due to quantum 
fluctuations.  So how large
are those fluctuations?  I will briefly discuss (in section 2)
a gedankan experiment to 
measure a distance $l$ and
deduce the intrinsic limitation $\delta l$
to the accuracy with which we can measure that 
distance, \cite{ng94,ng95,Karol} for that
distance undergoes quantum fluctuations.  
\footnote{I will further show that the scaling of 
$\delta l \stackrel{>}{\sim} l^{1/3} l_P^{2/3}$, as deduced from the 
gedankan experiment, is exactly what the holographic principle 
\cite{holography,Susskind} demands, according to which the maximum 
amount of information stored in a region of space scales as the area
of its two-dimensional surface, like a hologram.\cite{ng02}}
To gain more insight I will use (in section 2) an argument 
in mapping out the geometry of spacetime \cite{llo04} to arrive
at the same result.  When I generalize the argument to the case of an 
expanding Universe, we will
see that something akin to a (positive) cosmological constant emerges  
\cite{ng08} --- an 
effective dynamical cosmological
constant that has its origin in the quantum fluctuations of spacetime.  This 
dynamical cosmological constant
will be shown \cite{PRL,ng03} (in section 3)
to have the same magnitude as the one deduced by using 
unimodular gravity \cite{Bij,ht} in
combination with causal-set theory \cite{sorkin}.\\

Next I switch gear to discuss (in section 4)
gravitational thermodynamics /entropic gravity, inspired by the work of Ted 
Jacobson \cite{Jacob95} and Eric Verlinde \cite{verlinde}.  
By generalizing their work to our spacetime with positive 
cosmological constant $\Lambda$,
we will be led to a critical acceleration parameter $a_c$ of the same 
magnitude as the one introduced by
Milgrom by hand in his formulation of MOND (modified Newtonian dynamics) to 
explain flat
galactic rotation curves.  But I will argue that $a_c$ actually is a 
manifestation of the existence of dark
matter of a specific mass profile.  My collaborators and I call that model of 
dark matter ``modified dark
matter" (MDM) to distinguish it from cold dark matter (CDM). \cite{HMN,PRD} 
Recently my collaborators
and I have sucessfuly tested MDM (see Section 5) with galactic rotation curves
and galactic clusters. \cite{Edm1}\\

The take-home message from this talk is this:  It is possible that the dark 
sector (viz., dark energy and dark matter) has its origin in quantum gravity. 
If so, then we can perhaps understand why the dark sector is really so 
different from ordinary matter.  And if the scenario to be sketched in Section
6 is correct, then we can expect some rather novel particle phenomenology, 
for the quanta of the dark sector obey not the familiar Bose-Einstein or
Fermi-Dirac statistics, but an exotic statistics that goes by the name
infinite statistics \cite{DHR,govorkov,greenberg,fredenhagen}
or quantum Boltzmann statistics. \cite{plb,PRD}  However, it is known
that theories of particles obeying this exotic statistics are non-local ---
meaning that we cannot use conventional quantum field theories to describe
these particles' interactions.  On the positive side, this non-locality
may explain why so far dark matter detection experiments have failed 
to definitively detect dark matter.  Furthermore we expect that
the extended nature of the quanta of the dark sector may
connect them to certain global aspects of spacetime such as the cosmological
constant and the Hubble parameter (as will be shown in Section 4).
\\

I would like to take this opportunity to make a disclaimer on my own
behalf: In a recent paper
``New Constraints on Quantum Gravity from X-ray and Gamma-Ray
Observations" by Perlman, Rappaport, Christiansen, Ng, DeVore, and
D. Pooley \cite{perlman15}, it was claimed that
detections of quasars at
TeV energies with ground-based Cherenkov
telescopes seem to have ruled out the holographic spacetime foam model
(with $\delta l$ scaling as $l^{1/3} l_P^{2/3}$).  But now I believe 
this conclusion is conceivably premature when correct averaging is carried out.
The point is that these authors (including myself!) have considered the 
instantaneous fluctuations in the distance
between the location of the emission and a given point on the telescope 
aperture.
Perhaps one should average over both the huge number of Planck timescales
during
the time it takes light to propagate through the telescope system, and over
the
equally large number of Planck squares across the detector aperture. It is
then possible that the net fluctuations are exceedingly small, but
at the moment there is no formalism for carrying out such averages.
\cite{perlman16} \\

\section{\bf {Spacetime (Quantum) Foam and Effective Cosmological Constant 
$\Lambda$}}

Spacetime is foamy due to quantum fluctuations.
To examine how
large the fluctuations are, let us consider a gedankan experiment
in which a light signal is sent from a clock to a mirror (at a
distance $l$ away) and back to the clock in a timing experiment to
measure  $l$.  From the jiggling of the clock's position alone, the Heisenberg
uncertainty principle yields $\delta l^2 \stackrel{>}{\sim}
\frac{\hbar l}{mc}$, where $m$ is the mass of the clock. On the
other hand, the clock must be large enough not to collapse into a
black hole; this requires $\delta l \stackrel{>}{\sim} \frac{Gm}{c^2}$.  We
conclude that the fluctuations of a distance $l$ scales as
\begin{equation}
\delta l \stackrel{>}{\sim} l^{1/3} l_P^{2/3}, 
\label{deltal}
\end{equation}
where $l_P  = \sqrt{\hbar G/ c^3} \sim 10^{-33}$cm is
the Planck length. \footnote{Now
the amount of fluctuations in the distance
$l$ can be thought of as an accumulation of the
$l/l_P$ individual fluctuations each by an amount plus or minus $l_P$.
But note that the individual fluctuations cannot be completely
random (as opposed to random-walk); actually
successive fluctuations must be {\it entangled} and somewhat
{\it anti-correlated}
(i.e., a plus fluctuation is slightly more likely followed by a minus
fluctuation and vice versa),
in order that together they produce a total fluctuation less
than that in a random-walk model (for which 
$\delta l \stackrel{>}{\sim} l^{1/2} l_P^{1/2}$.) \cite{yjng05}
This small amount of
anti-correlation between successive fluctuations (corresponding to
what statisticians call fractional Brownian motion with
self-similarity parameter $\frac{1}{3}$)
must be due to quantum gravity effects.}
\cite{ng94,ng95,Karol}\\

One can further show that the scaling of $\delta l$ given above is exactly 
what
the holographic principle \cite{holography,Susskind} demands.
Heuristically, this comes about because a cube with side $l$ contains $\sim
l^2/l_P^2$ number
of small cubes with side $\delta l$. \cite{ngvd00}
Imagine partitioning a cubic region with side $l$ into small cubes. 
The small cubes so constructed should
be as small as physical laws allow so that intuitively we can associate
one degree of
freedom with each small cube.   In other words, the number of
degrees of freedom that the region can hold is given by the number of
small
cubes that can be put inside that region.
A moment's thought tells us that each side of a small cube
cannot be smaller than the accuracy
$\delta l$ with which we can measure each side $l$ of the big cube.
This can be easily shown by applying the method of contradiction:  assume
that we can construct small cubes each of which has sides less than
$\delta l$.  
Then by lining up a row of such small cubes along a side of  
the big cube from end to end, and by counting the number of such small
cubes,
we would be able
to measure that side (of length $l$) of the big cube
to a better accuracy than $\delta l$.  But, by
definition, $\delta l$ is the best accuracy with which we can measure
$l$.  The
ensuing contradiction is evaded by the realization that each of the
smallest
cubes (that can be put inside the big cube)
indeed measures $\delta l$ by
$\delta l$ by
$\delta l$.  Thus, the number of degrees of freedom $I$ in the region
(measuring $l$ by $l$ by $l$) is given by $l^3 / \delta l^3$,
which,
according to the holographic principle, is  
\begin{equation}
I \stackrel{<}{\sim} l^2 / l_p^2.
\end{equation}
\label{holog}
It follows
that $\delta l$  is bounded (from below) by the cube root of $l l_P^2$,
the same result as found
above in the gedanken experiment argument.\\

We can rederive the scaling of $\delta l$ by another argument. 
Let us consider mapping
out the geometry of spacetime for a spherical volume of radius $l$ over the
amount of time $2l/c$ it takes light to cross the volume.\cite{llo04} One way to 
do this is to fill the space with clocks, exchanging
signals with the other clocks and measuring the signals' times of arrival. The 
total number of operations, including the ticks of the clocks and
the measurements of signals, is bounded by the Margolus-Levitin
theorem \cite{mar98}
which stipulates that the rate of operations cannot exceed the amount of energy 
$E$ that is available for the operation divided by $\pi \hbar/2$.  This theorem, 
combined with the
bound on the total mass of the clocks to prevent black hole formation, implies 
that the total number of operations that can occur in this spacetime volume is 
no bigger than $2 (l/l_P)^2 / \pi$.  To maximize spatial resolution,
each clock must tick
only once during the entire time period.  If we regard the operations
as partitioning the spacetime volume into ``cells", then on the average each 
cell
occupies a spatial volume no less than $\sim l^3 / (l^2 / l_P^2)
= l l_P^2$, yielding an average separation between neighboring
cells no less than $ \sim l^{1/3} l_P^{2/3}$.
\cite{ng08}  This spatial separation can be interpreted as the average minimum 
uncertainty in the
measurement of a distance $l$, that is, $\delta l \stackrel{>}{\sim} l^{1/3}
l_P^{2/3}$, 
in agreement with the result found in the gedanken 
experiment to measure the fluctuation of a distance $l$.\\

We make two observations: \cite{Arzano,plb} First, maximal
spatial resolution (corresponding to  $\delta l \sim l^{1/3}
l_P^{2/3}$)
is possible only if the maximum energy density $\rho \sim (l l_P)^{-2}$
is available to map the geometry
of the spacetime region, without causing a gravitational collapse.
Secondly,
since, on the average, each cell occupies a spatial volume of $l l_P^2$,
a spatial region of size $l$ can contain no more than $\sim l^3/(l l_P^2) = 
(l/l_P)^2$ cells.
Hence, this result for spacetime fluctuations 
corresponds to the case of
maximum number of bits of information $l^2 /l_P^2$
in a spatial region of size $l$, that is
allowed by the holographic principle\cite{holography,Susskind}.\\

It is straightforward to generalize \cite{Arzano} the above discussion for a 
static spacetime region with low spatial curvature to the
case of an expanding
universe by the substitution of $l$ by $H^{-1}$ in the expressions for
energy and entropy densities, where $H$ is the Hubble parameter. (Henceforth we 
adopt $c=1=\hbar$ for convenience unless stated otherwise.)  Thus,
applied to cosmology, the above argument leads to the prediction
that (1) the cosmic energy density has the critical value 
\begin{equation}
\rho \sim (H/l_P)^2, 
\end{equation}
and (2) the universe
of Hubble size $R_H$ contains $I \sim (R_H/l_p)^2$ bits of 
information. (For the present cosmic epoch we have $I \sim 10^{122}$.)
It follows that the average energy carried by each particle/bit is
$\rho R_H^3/I \sim R_H^{-1}$.
Such long-wavelength constituents of dark energy give rise to
a more or less uniformly distributed cosmic energy density and
act as a dynamical cosmological constant with the observed small but nonzero 
value
\begin{equation}
\Lambda \sim 3 H^2.\\
\label{cc}
\end{equation}

\section{\bf Cosmological Constant $\Lambda$ via
Unimodular Gravity and Causal-set Theory}
The dynamical cosmological constant we have just obtained will be seen
to play an important role in our subsequent discussions.  So let us 
``rederive" it by using another method based on quantum gravity.
The idea makes use of the theory of
unimodular gravity\cite{Bij,PRL} (which
can be regarded as the ordinary theory of gravity except for the
way the cosmological constant $\Lambda$ arises in the theory).  But here
we will use the (generalized) version of unimodular gravity given by the
Henneaux and Teitelboim action\cite{ht}
\begin{equation}
S_{unimod} = - \frac{1}{16 \pi G} \int [ \sqrt{g} (R + 2 \Lambda) - 2
\Lambda
\partial_\mu {\mathcal T}^\mu](d^3x)dt.
\label{HT}
\end{equation}
One of its equations of motion is $\sqrt{g} = \partial_\mu
\mathcal{T}^\mu$,
the generalized unimodular condition, with $g$ given in terms of the
auxiliary field $\mathcal{T}^{\mu}$.  Note that, in this theory,
$\Lambda / G$ plays the role of
``momentum" conjugate to the ``coordinate" $\int d^3x {\mathcal T}_0$ which
can
be identified, with the aid of the generalized unimodular condition, as
the spacetime volume $V$.  Hence $\Lambda /G$ and $V$ are
conjugate to each other.  It follows that
their fluctuations obey a Heisenberg-type
quantum uncertainty principle,
\begin{equation}
\delta V \! \delta \Lambda/G \sim 1.
\label{heisenb}
\end{equation}

Next we borrow an argument due to Sorkin\cite{sorkin},
drawn from the causal-set theory, which
stipulates that continous geometries in classical gravity should be
replaced by ``causal-sets", the discrete substratum of spacetime.
In the framework of the causal-set theory, the
fluctuation in the number of elements $N$ making up the set is of the
Poisson type, i.e., $\delta N \sim \sqrt{N}$.  For a causal set, the
spacetime volume $V$ becomes $l_P^4 N$.  It follows that
\begin{equation}
\delta V \sim l_P^4\delta N \sim l_P^4 \sqrt{N}
\sim l_P^2\sqrt{V} = G \sqrt{V}.
\label{poisson}
\end{equation}
Putting Eqs. (\ref{heisenb}) and (\ref{poisson}) together yields a minimum
uncertainty in
$\Lambda$\ of $\delta \Lambda \sim V^{-1/2}$. This cosmological constant,
like the one given by Eq. (\ref{cc}) from a heuristic quantum mechanical
consideration, is finite and is to be identified with the fully 
renormalized cosmological constant from a quantum field-theoretic
argument given by the path integration method, to which we turn next.\\

Following an argument due to Baum\cite{Baum}, Hawking\cite{Hawk}, 
and Adler\cite{Adler}, 
one can now plausibly argue \cite{PRL} that, in the
framework of unimodular gravity,
$\Lambda$ vanishes to the lowest order of
approximation and that it is positive if it is not zero. 
The argument goes as follows: Consider
the vacuum
functional for unimodular gravity given by path integrations over
$\mathcal{T}^{\mu}$, $g_{\mu \nu}$, the matter fields (represented
by $\phi$), and $\Lambda$:
\begin{equation}
Z_{Minkowski} = \int d\mu (\Lambda) \int d [\phi] d [g_{\mu \nu}] \int d
[{\mathcal T}^{\mu}]  exp \left\{ -i[ S_{unimod} + S_{M}(\phi, g_{\mu
\nu})]\right\}, 
\label{Z}
\end{equation}
where $S_{M}$ stands for the contribution from matter (including 
radiation) fields (and $d \mu
(\Lambda)$ denotes the measure of the $\Lambda$ integration). 
\cite{Adler} The
integration over $\mathcal{T}^{\mu}$ yields $\delta(\partial_{\mu}
\Lambda)$, which implies that $\Lambda$ is spacetime-independent (befiting
its role as the cosmological constant).  A Wick rotation now allows us to
study the Euclidean vacuum functional $Z$.  The integrations over $g_{\mu
\nu}$ and $\phi$ give $exp[-S_{\Lambda}(\overline{g}_{\mu \nu},
\overline{\phi})]$ where $\overline{g}_{\mu \nu}$ and $\overline{\phi}$
are the background fields which minimize the effective action
$S_{\Lambda}$.  A curvature expansion for $S_{\Lambda}$ yields a
Lagrangian whose first two terms are the Einstein-Hilbert terms $\sqrt{g}
(R + 2\Lambda)$.  Note that (1) 
$\Lambda$ now denotes the fully renormalized
cosmological constant after integrations over all other fields have 
been carried out; (2) the Einstein-Hilbert terms are exactly the first
two terms in Eq. (\ref{HT}), hence the fluctuation $\delta \Lambda \sim 
V^{-1/2}$ we found above now applies to the renormalized $\Lambda$.
Next we can make a change of variable from the original
(bare) $\Lambda$ to the renormalized $\Lambda$ 
for the integration in Eq. (\ref{Z}).  Let us assume that for
the present and recent cosmic eras, 
$\phi$ is essentially in the ground state, then
it is reasonable to neglect the effects of $\overline{\phi}$. 
\cite{ngvd01}.  To
continue, we follow Baum\cite{Baum} and Hawking\cite{Hawk} to evaluate
$S_{\Lambda}(\overline{g}_{\mu \nu}, 0)$.  For negative $\Lambda$,
$S_{\Lambda}$ is positive; for positive $\Lambda$, one finds
$S_{\Lambda}(\overline{g}_{\mu \nu}, 0) = -3 \pi / G\Lambda$, so that 
\begin{equation}
Z \approx \int \! d\mu (\Lambda) exp(3 \pi /G \Lambda).
\label{finalZ}
\end{equation}
This implies that the observed cosmological constant in
the present and recent eras is essentially zero 
(or more accurately, very small but positive).  So we \cite{PRL}
conclude
that $\Lambda$ is positive and it fluctuates about zero with a magnitude of
\begin{equation}
\Lambda \sim V^{-1/2} \sim R_H^{-2}, 
\end{equation}
where, we recall, $R_H$ is the Hubble radius of the Universe,
contributing an energy density $\rho$ given by:
$
\rho \! \sim \! + \frac{1}{l_P^2 R_H^2},
$
which is of the order of the critical density as observed!\\

\section{\bf {From Cosmological Constant $\Lambda$ to Modified Dark Matter 
(MDM)}}

The dynamical cosmological constant (originated from quantum fluctuations of 
spacetime) can now be
shown to give rise to a critical acceleration
parameter in galactic dynamics.  The argument \cite{HMN} is based on
a simple generalization of E. Verlinde's recent proposal of entropic gravity 
\cite{verlinde,Jacob95} for $\Lambda = 0$ to the case of de-Sitter space 
with positive $\Lambda$.  Let us first review Verlinde's derivation (or
prescription, if you like) of Newton's second law $\vec{F} = m \vec{a}$.  
Consider a particle with mass $m$ approaching a holographic screen
at temperature $T$.  Using the first law of thermodynamics to introduce the 
concept of entropic force
$
F = T \frac{\Delta S}{\Delta x},
$
and invoking Bekenstein's original arguments \cite{bekenstein}
concerning the entropy $S$ of black holes,
$\Delta S = 2\pi k_B \frac{mc}{\hbar} \Delta x$,
Verlinde gets $ F = 2\pi k_B \frac{mc}{\hbar} T$.  With the aid of
the formula for the Unruh temperature, $k_B T = \frac{\hbar a}{2 \pi c}$,
associated with a uniformly accelerating (Rindler) observer, Verlinde
then obtains $\vec{F} = m \vec{a}$.
Now in a de-Sitter space with cosmological 
constant $\Lambda$, the net Unruh-Hawking temperature, 
\cite{unruh,Davies,hawking} as measured by a non-inertial observer with 
acceleration $a$ relative to an inertial observer, is 
\begin{equation}
\tilde{T} = \frac{\hbar \tilde{a}}{2\pi k_B c}, 
\end{equation}
with \cite{deser}
\begin{equation}
\tilde{a} = \sqrt{a^2+a_0^2} - a_0, 
\end{equation}
where $a_0 \equiv \sqrt{\Lambda / 3}$.  Hence the
entropic force (in de-Sitter space) is given by the replacement of $T$ and 
$a$ by  $\tilde{T}$ and $\tilde{a}$ respectively, leading to
\begin{equation}
F =  m [\sqrt{a^2+a_0^2}-a_0].
\end{equation}
For $ a \gg a_0$, we have $F/m \approx a$ which gives $a = a_N \equiv GM/r^2$, 
the familiar Newtonian value for the acceleration due to the source $M$. But for 
$a \ll a_0$, $F \approx m \frac{a^2}{2\,a_0},$ so
the terminal velocity $v$ of the test mass $m$ in a circular motion with radius 
$r$ should be determined from
\,$ m a^2/(2a_0) = m v^2/r$.  In this small acceleration regime,
the observed flat galactic rotation curves ($v$ being independent of $r$) now 
require
$ a \approx \left(2  a_N \,a_0^3 \,/ \pi \right)^{\frac14}$.
But that means $F \approx m \sqrt{a_N a_c}\,$.
This is the celebrated modified Newtonian dynamics (MoND)
scaling \cite{mond,FandM,interpol} discovered by Milgrom who introduced the 
critical acceleration parameter 
\begin{equation}
a_c = a_0/ (2 \pi) = c H / (2 \pi) 
\end{equation}
by hand to phenomenologically explain the 
observed flat galactic rotation curves.  
Thus, we have recovered MoND with the correct
magnitude for the critical galactic acceleration parameter $a_c \sim 
10^{-8} cm/s^2$.
From our perspective, MoND is a {\it classical} phenomenological
consequence of {\it quantum} gravity (with the $\hbar$ dependence
in $T \propto \hbar$ and $S \propto 1/\hbar$ cancelled out in the 
product $TS$ for the entropic force). \cite{HMN}  As a 
bonus, we have also recovered the observed Tully-Fisher relation ($v^4 \propto 
M$).\\

Having generalized Newton's 2nd law, we \cite{HMN}
can now follow the second half of Verlinde's argument \cite{verlinde} to 
generalize Newton's law of gravity
$a= G M /r^2$.  Verlinde derives Newton's law of gravity by considering an
imaginary quasi-local (spherical) holographic screen of area $A=4 \pi
r^2$ with temperature $T$, and by invoking the equipartition of energy $E= 
\frac{1}{2} N k_B T$
with $N = Ac^3/(G \hbar)$ being the total number of degrees of freedom (bits) on 
the screen, as well as the Unruh
temperature formula  $k_B T = \frac{\hbar a}{2 \pi c}$,
and the fact that $E= M c^2$.  The generalized Newton's
law of gravity (for the case of de-Sitter space) is obtained by the 
replacement of $T$ and $M$ by $\tilde{T}$ and $\tilde{M}$ respectively, so 
that we get 
\begin{equation}
2 \pi k_B \tilde{T} = G\,\tilde{M} /r^2,
\end{equation}
where 
\begin{equation}
\tilde{M} = M + M_d 
\end{equation}
represents the \emph{total} mass enclosed within the 
volume $V = 4 \pi r^3 / 3$, with
$M_d$ being some unknown mass, i.e., dark
matter.  For $a \gg a_0$, consistency with the Newtonian force law $a \approx 
a_N$ implies $M_d \approx 0$.  But
for $a \ll a_0$, consistency with the condition 
$a \approx \left(2 a_N \,a_0^3 /\pi \right)^{\frac14}$ 
requires \footnote{Actually the two acceleration limits have little to 
say about the intermediate regime; thus we expect that a more generic 
dark mass profile is of the form 
$
M_d = \left[\,\xi\,\left(\,\frac{a_0}{a}\,\right)+\frac{1}{\pi}\,
\left(\,\frac{a_0}{a}\,\right)^2\,\right] \, M\,
$  
with positive parameter $\xi \sim 1$.  See discussions in the next section
about the dark matter mass profile.}
\begin{equation}
M_d \approx \frac{1}{\pi} \left(\,\frac{a_0}{a}\,\right)^2\, M
\sim (\sqrt{\Lambda}/G)^{1/2}M^{1/2}r.  
\label{darkmatt}
\end{equation}
This yields
the dark matter mass density $\rho_d$ profile given by
$
\rho_d(r) \sim M^{1/2}(r_v) (\sqrt{\Lambda}/G)^{1/2} / r^2,
$ for an ordinary (visible) matter source of radius $r_v$ with
total mass $M(r_v)$. \footnote{
This result can be compared with the distribution associated with
an isothermal Newtonian sphere in hydrostatic equilibrium (used by some dark 
matter proponents):
$
\rho (r) = \sigma (r^2 + r_0^2)^{-1}.
$
Asymptotically the two expressions agree with
$\sigma$ identified as $\sim M^{1/2}(r_v) (\sqrt{\Lambda}
/G)^{1/2} $. }\\

Thus dark matter indeed exists!
And the MoNDian force law derived above, at the galactic scale, is 
simply a manifestation of dark matter! \cite{HMN,turnerkip}
Dark matter of this kind can behave as if there is no dark
matter but MoND. Therefore, we used to call it ``MoNDian dark matter" 
which, to some people sounds like an oxymoron.  Now we call it
``modified dark matter".
Note that the dark matter profile we have obtained relates, at the 
galactic scale,
dark matter ($M_d$), dark energy ($\Lambda$) and ordinary matter
($M$) to one another.\\

\section{\bf Observational Tests of MDM}

In order to test MDM with galactic rotation curves, we fit computed rotation 
curves to a selected sample of Ursa Major galaxies given in \cite{Sanders98}. 
The sample contains both high surface brightness (HSB) and low surface 
brightness (LSB) galaxies. The rotation curves, predicted by MDM 
as given above by 

\begin{equation}
F =  m [\sqrt{a^2+a_0^2}-a_0]
  = m a_N \left[ 1 + \frac{1}{\pi} \left( \frac{a_0}{a} \right)^2 \right],
\end{equation}
along with $ F = m v^2 / r$ for circular orbits, can be
solved for $a(r)$ and $v(r)$. We \cite{Edm1}
fit these to the observed rotation curves as determined in \cite{Sanders98}, 
using a least-squares fitting routine. As in \cite{Sanders98}, the mass-to-light 
ratio $M/L$, which is our {\it only} fitting parameter for MDM, is assumed 
constant for a given galaxy but allowed to vary between galaxies. Once we have 
$a(r)$, we can find the MDM density profile by using $M_d \approx 
\frac{1}{\pi} 
\left(\,\frac{a_0}{a}\,\right)^2\, M$
to give
$
\rho_d (r) \;=\; \left( \frac{a_c}{r} \right)^2 \frac{d}{dr} \left( 
\frac{M}{a^2} 
\right).
$
\\

\begin{figure}[h]
\hspace{2pc}
\begin{minipage}{15pc}
\includegraphics[angle=90,width=15pc]{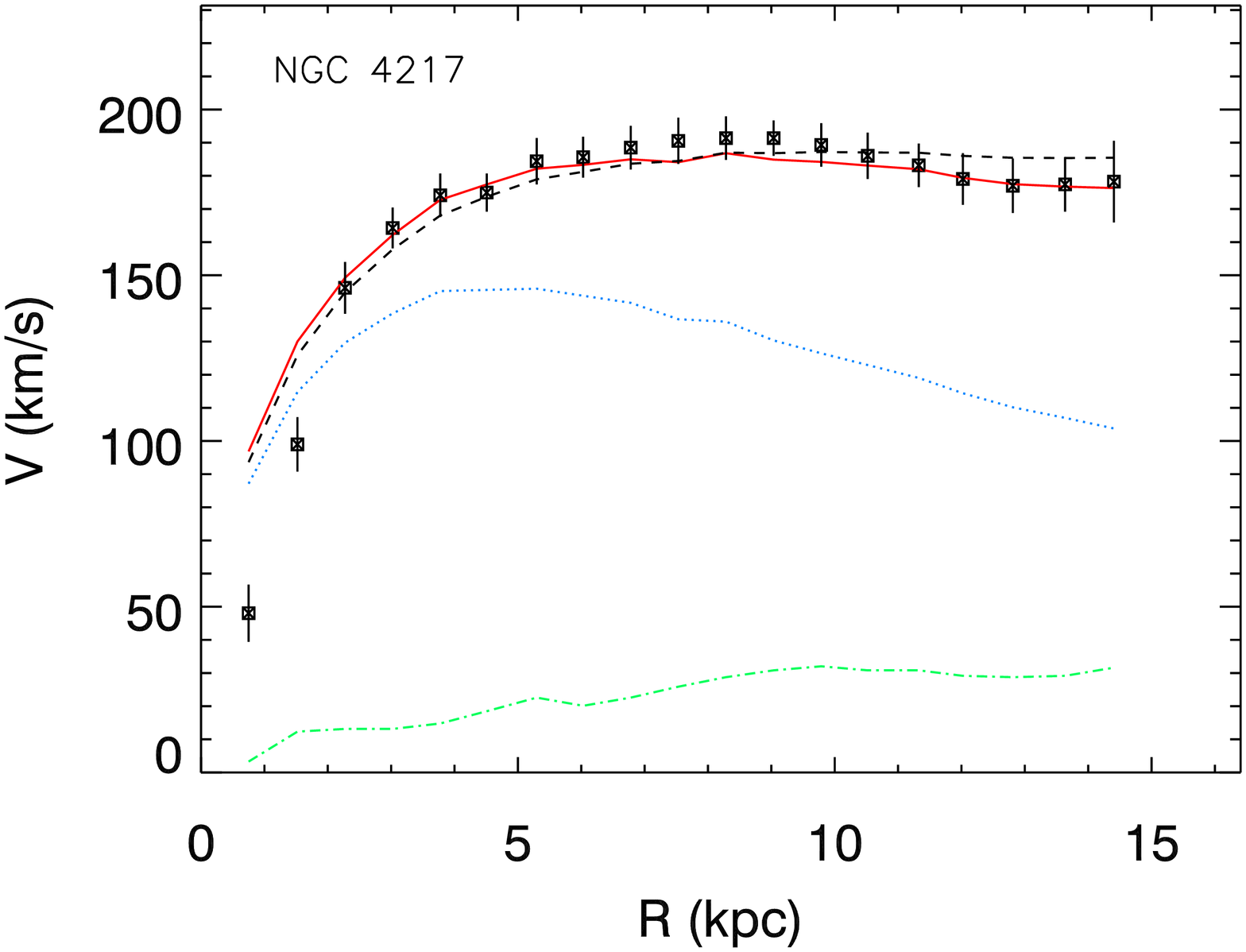}
\caption{\label{Fig1}Galactic rotation curves for NGC 4217 (HSB).}
\end{minipage}\hspace{4pc}%
\begin{minipage}{15pc}
\includegraphics[angle=90,width=15pc]{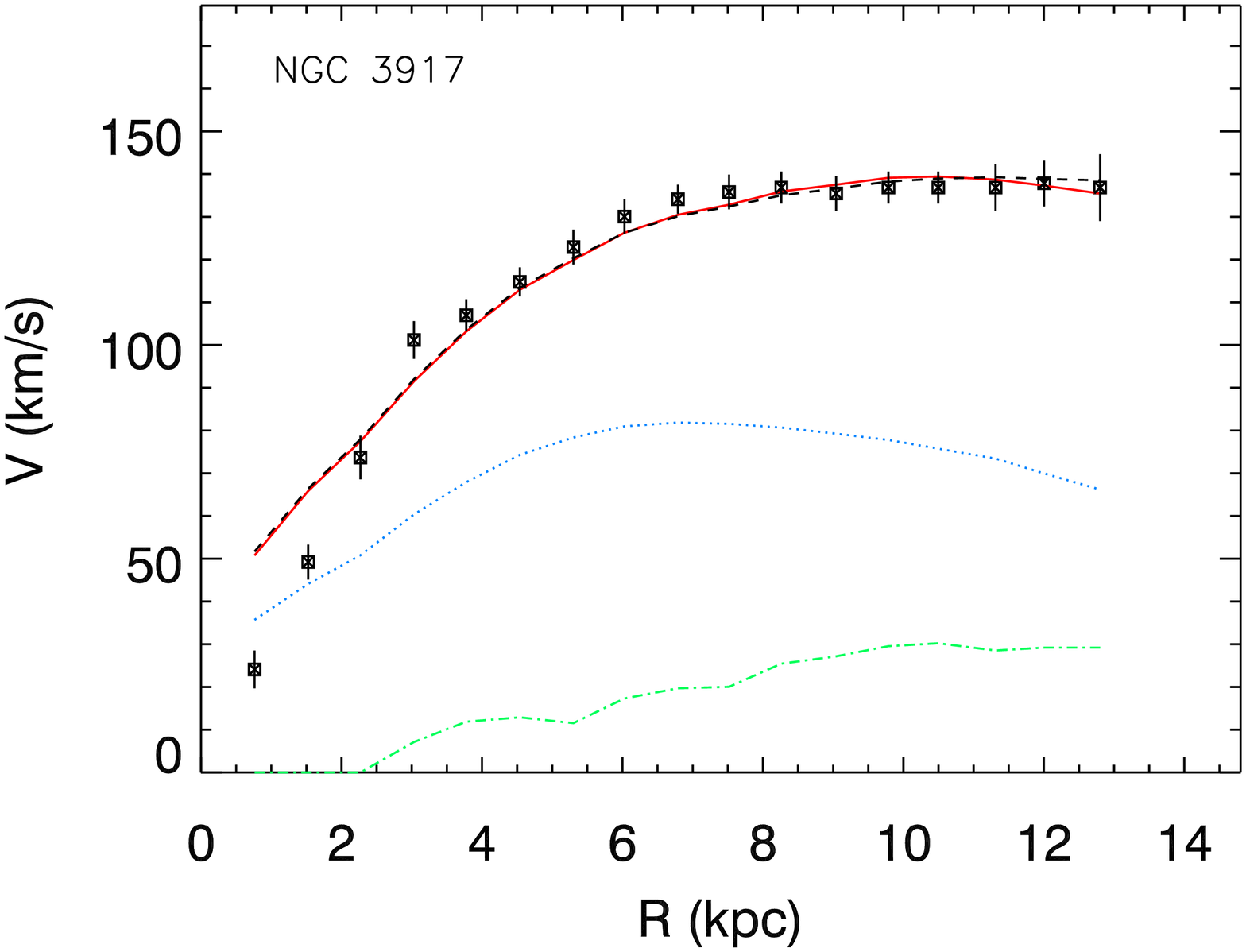}
\caption{\label{Fig2}Galactic rotation curves for NGC 3917 (LSB).}
\end{minipage} \end{figure}

Rotation curves predicted by MDM for NGC 4217, a typical HSB galaxy, and NGC 
3917, a typical LSB galaxy in the sample
are shown in Fig. ~\ref{Fig1} and Fig. ~\ref{Fig2} respectively. 
(See Ref. \cite{Edm1} for the rotation curves for the other 28 galaxies.)


\vspace{0.5cm}

 In these figures, observed rotation curves are depicted as filled circles with 
error bars, and for the two curves at the bottom, the dotted and dash-dotted 
lines show the stellar and interstellar gas rotation curves, respectively. The 
solid lines and dashed lines are rotation curves predicted by MDM and the 
standard cold dark matter (CDM) paradigm respectively. For the CDM fits, we use 
the Navarro, Frenk \& White (NFW) \cite{nfw} density profile, employing {\it 
three} free parameters (one of which is
the mass-to-light ratio.)  It is fair
to say that both models fit the data well; \footnote{We should point out that
the rotation curves predicted by MDM and MOND have been found \cite{Edm1} to be 
virtually indistinguishable
over the range of observed radii and both employ only 1 free parameter.}
but while the MDM fits use only 1 free parameter,
for the CDM fits one needs to use 3 free parameters.  Thus the MDM model is a 
more economical model than CDM in fitting data at the galactic scale.\\

\begin{figure}[h]
\hspace{2pc}
\begin{minipage}{15pc}
\includegraphics[angle=90,width=15pc]{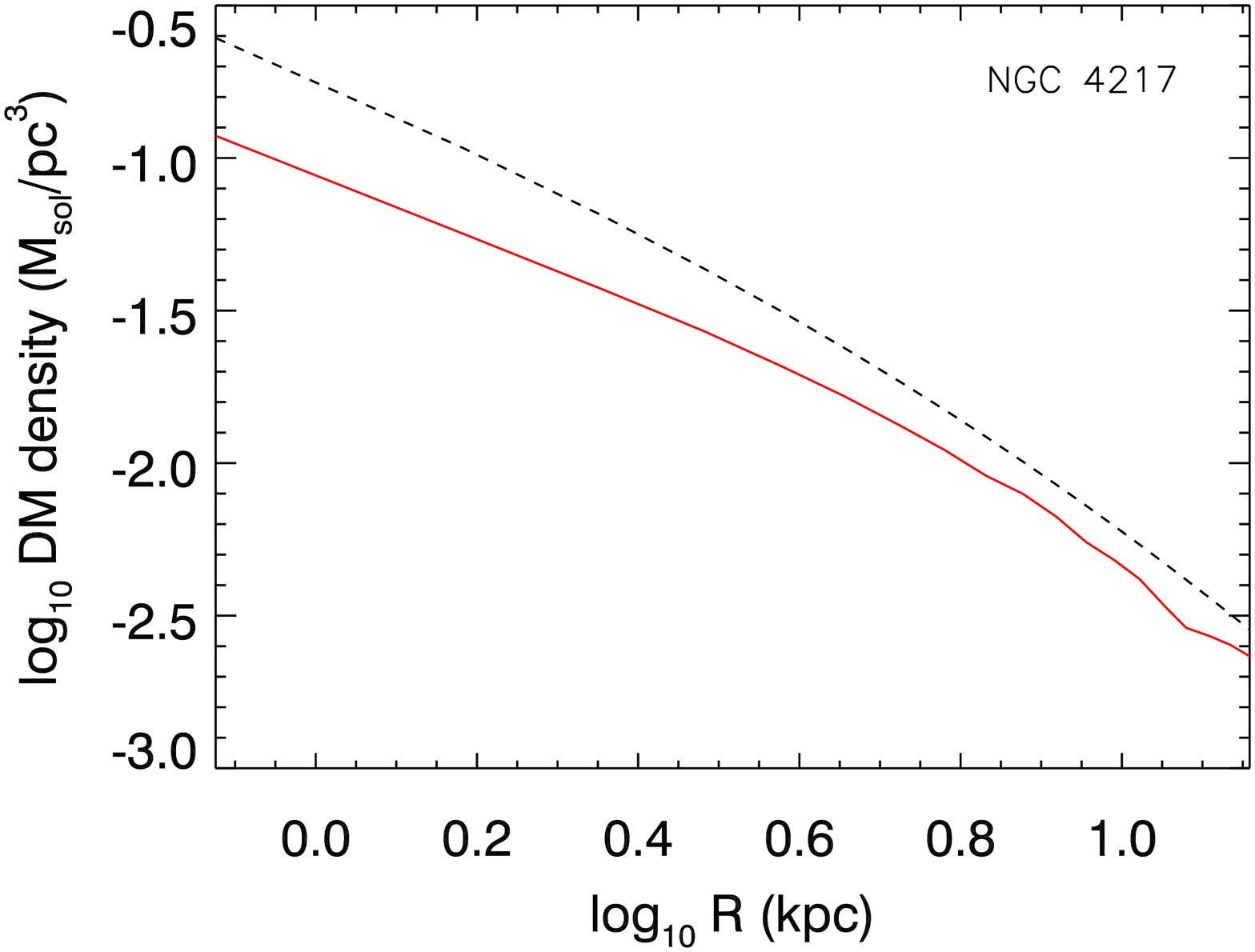}
\caption{\label{Fig3}Dark matter density profile for NGC 4217 (HSB).}
\end{minipage}\hspace{4pc}%
\begin{minipage}{15pc}
\includegraphics[angle=90,width=15pc]{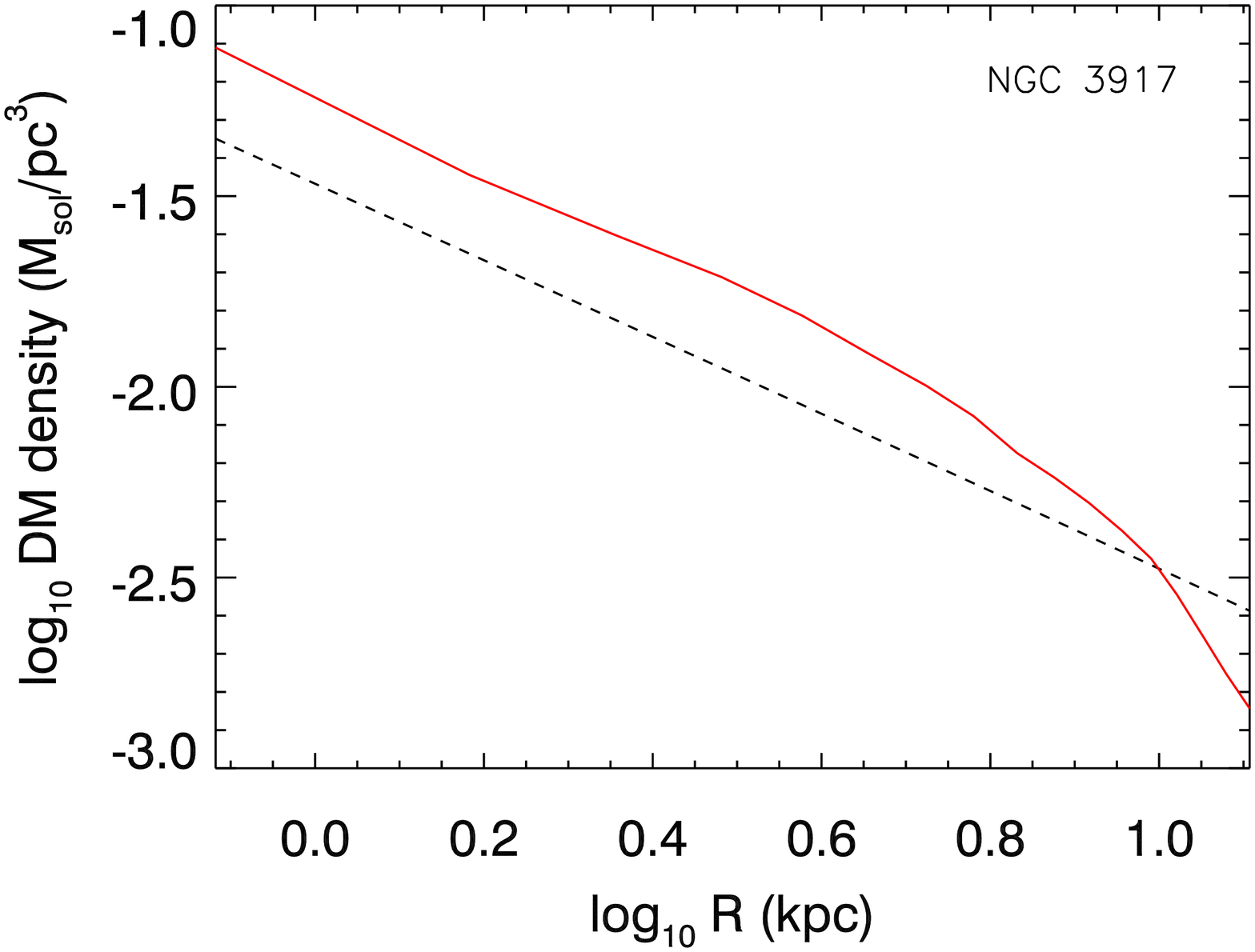}
\caption{\label{Fig4}Dark matter density profile for NGC 3917 (LSB).}
\end{minipage} \end{figure}

Shown in Fig. ~\ref{Fig3} and Fig. ~\ref{Fig4} are the dark matter 
density
profiles predicted by MDM (solid lines) and CDM (dashed lines) for
the HSB galaxy NGC 4217 and the LSB galaxy NGC 3917 in the sample 
respectively.
(See Ref. \cite{Edm1} for details.)\\

To test MDM with astronomical observations at a larger scale, we 
compare dynamical and observed masses in a large sample of galactic clusters. 
\footnote{The comparison is made in some unpublished work by D. Edmonds et al. 
\cite{Edm1}}
First, let us recall that
the MDM profile $M_d = \frac{1}{\pi}\,\left(\,\frac{a_0}{a}\,\right)^2\, M\,$
reproduces the flat rotation curves.  But
we expect that a more general profile should be of
the form
$
M_d = \left[\,\xi\,\left(\,\frac{a_0}{a}\,\right)+\frac{1}{\pi}\,
\left(\,\frac{a_0}{a}\,\right)^2\,\right] \, M\,,
$
with $\xi >0$ which ensures that $M_d > 0$ when $a \gg a_0$.
For the more general profile, the entropic force expression
is replaced by
\begin{equation}
F = m a_N \left[ 1 + \xi\,\left(\,\frac{a_0}{a}\,\right)+
\frac{1}{\pi} \left( \frac{a_0}{a} \right)^2 \right].
\label{force3}
\end{equation}

Sanders \cite{Sanders1999} studied the virial discrepancy (i.e.,
the discrepancy between the observed mass and the dynamical mass) in the 
contexts of Newtonian dynamics and MOND.  We \cite{Edm1} have adapted his 
approach to the case of MDM. For his work,
Sanders considered 93 X-ray-emitting clusters from the compilation
by White, Jones, and Forman (WJF) \cite{FJW1997}.
He found the well-known discrepancy
between the Newtonian dynamical mass ($M_{\textrm{N} }$) and the observed mass 
($M_{\textrm{obs} }$):
$
\left \langle \, \frac{M_\textrm{N}}{ M_{\textrm{obs} }} \,\right \rangle 
\approx 4.4\,.
$
And for the sample clusters, he found $\langle M_{\textrm{MOND}} / 
M_{\textrm{obs}} \rangle \approx 2.1.$\\

\begin{figure}[h]
\begin{minipage}{16pc}
\includegraphics[angle=0,width=16pc]{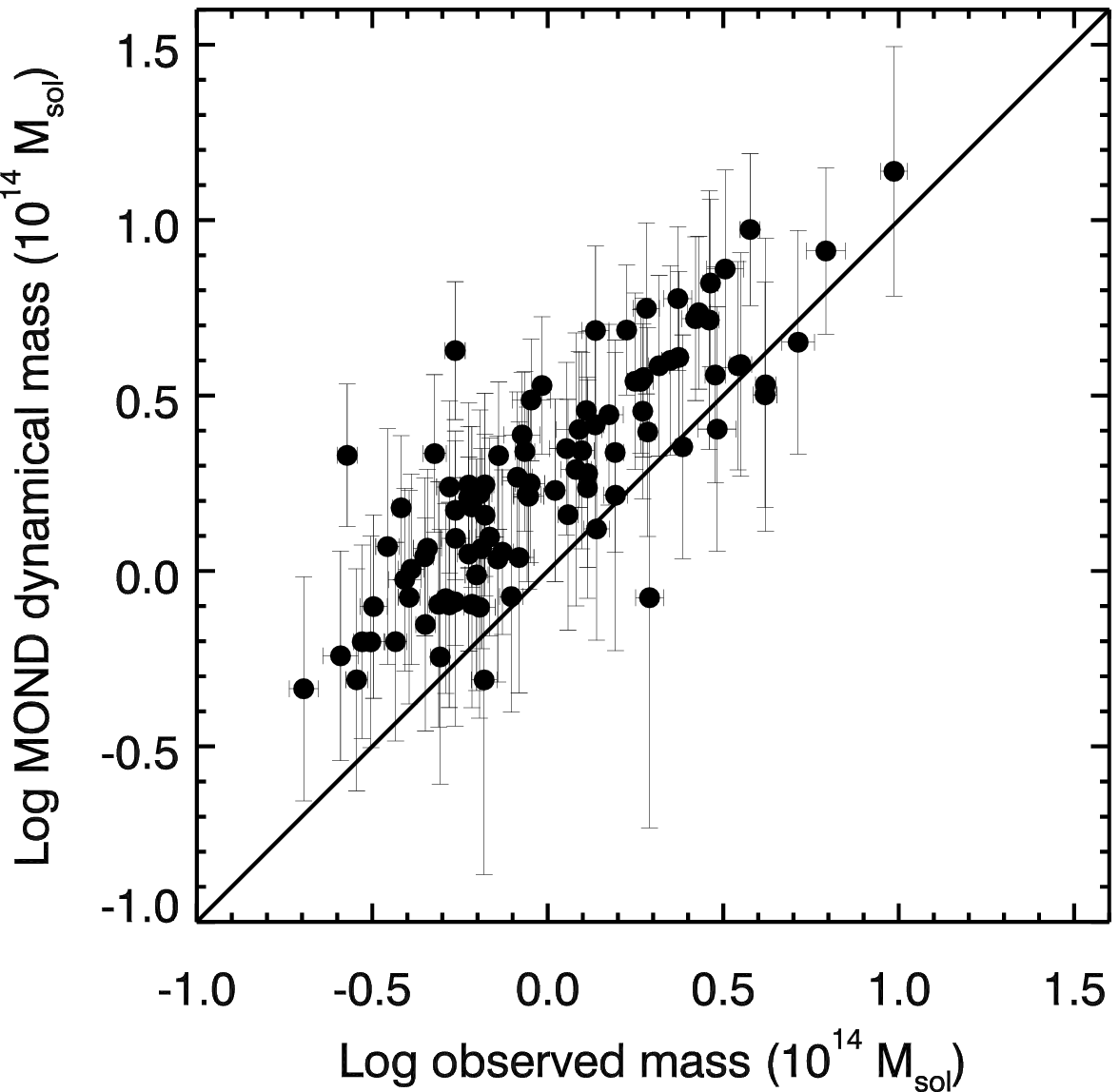}
\caption{\label{Fig5}Fit to galactic cluster data using
MONDian dynamics.}
\end{minipage}\hspace{4pc}%
\begin{minipage}{16pc}
\includegraphics[angle=0,width=16pc]{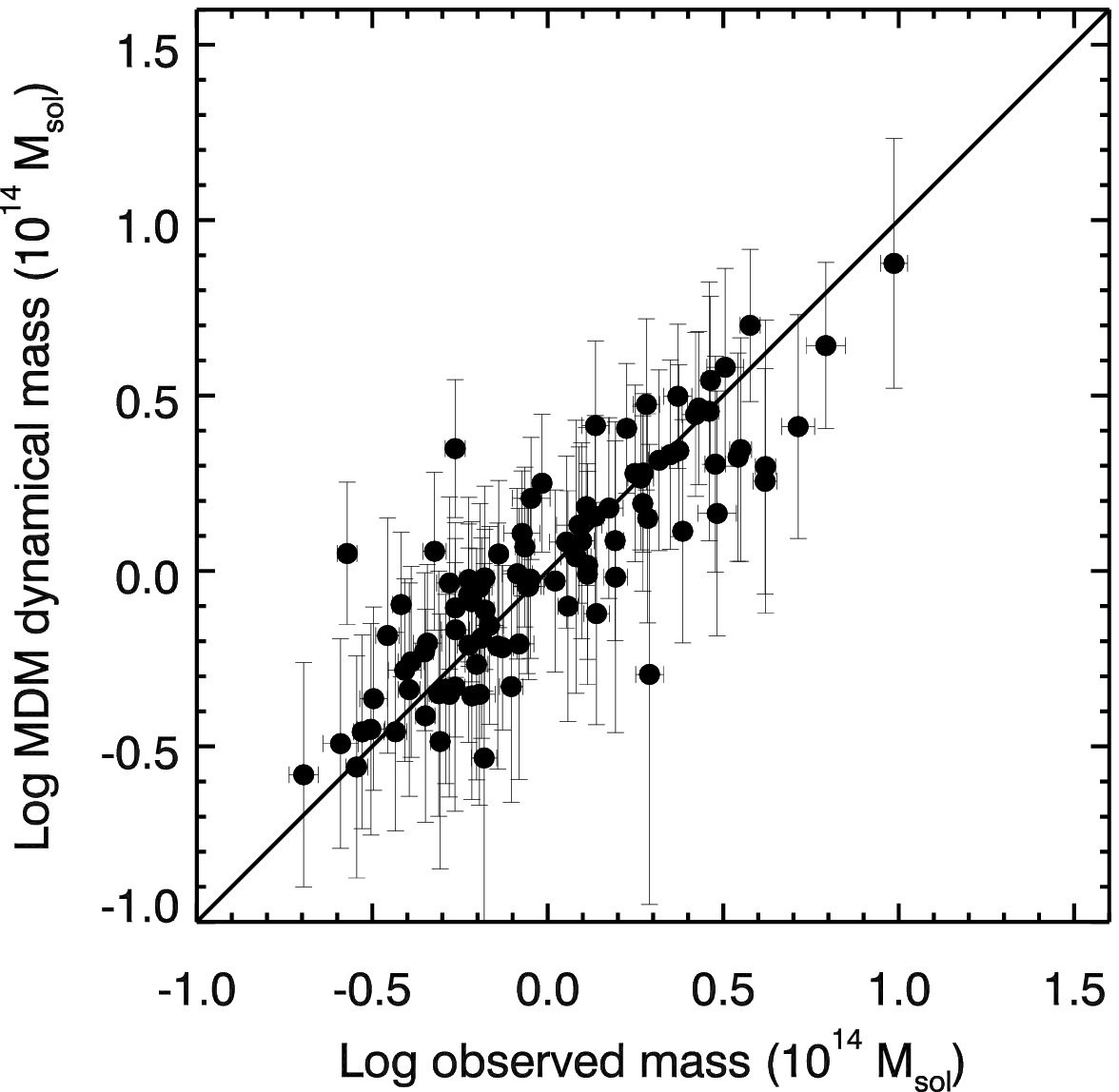}
\caption{\label{Fig6}Fit to galactic cluster data using
MDM dynamics.}
\end{minipage} \end{figure}

For MDM, the observed (effective) acceleration is given by
$a_{obs} = \sqrt{a^2 + a_0^2} - a_0$.  Using
the more general expression for the MDM profile,
we have $a_{obs} = \frac{G M_{MDM}}{r^2} \{ 1 + \xi\,\left(\,
\frac{a_0}{a}\,\right)+ \frac{1}{\pi}\,\left(\,\frac{a_0}{a}\,\right)^2 \}$.
Recalling that $a_{obs} = G M_N / r^2$ for Newtonian dyanmics, we get
\begin{equation}
M_{\textrm{MDM}} = \frac{M_\textrm{N}}{1 + \xi\,
\left(\,\frac{a_0}{a}\,\right)+ 
\frac{1}{\pi}\,\left(\,\frac{a_0}{a}\,\right)^2}\,,
\end{equation}
for the dynamical mass for MDM, using $\xi$ as a universal fitting parameter.
With $\xi \approx 0.5$, we get
$
\left \langle \, \frac{M_\textrm{MDM}}{ M_{\textrm{obs} }} \,\right \rangle
\approx 1.0\,.$
\footnote{
For completeness we mention that previously we
have used $\xi = 0$ when fitting galactic rotation curves.
But since now
the galaxy cluster sample in our current study implies $\xi \approx 0.5$, we
(in unpublished work \cite{Edm1})
refit the galaxy rotation curves using $\xi = 0.5$ and find the fits are 
nearly
identical with a reduction in mass-to-light ratios of about 35\%.
}
In Fig. ~\ref{Fig5} and Fig. ~\ref{Fig6}, we show the MOND and MDM 
dynamical masses respectively
against the total observed mass for the 93 sample clusters compiled by
WJF.
The virial discrepancy is eliminated in the context of MDM! Recalling 
that Sanders found
$\langle M_{\textrm{MOND}} / M_{\textrm{obs}} \rangle \approx 2.1$,
we conclude that, at the cluster 
scale, MDM is superior to MOND.\\

\section{\bf {The Dark Sector and Infinite Statistics}}

What is the essential difference between ordinary matter and
dark energy from our perspective?
To find that out, let us recall our discussions in Section 2, and liken
the quanta of dark energy to
a perfect gas of $N$ particles obeying Boltzmann statistics
at temperature $T$ in a volume $V$.  For the
problem at hand, as the lowest-order approximation, we can neglect the
contributions from matter and radiation to the cosmic 
energy density for the recent and present eras.  
Thus let us take $V \sim R_H^3$, $T \sim R_H^{-1}$, and $N \sim
(R_H/ l_P)^2$. A standard calculation (for the relativistic case) yields the
partition function $Z_N = (N!)^{-1} (V / \lambda^3)^N$, where
$\lambda = (\pi)^{2/3} /T$.
With the free energy given by
$F = -T ln Z_N = -N T [ ln (V/ N \lambda^3) + 1]$,
we get, for the entropy of the system,
\begin{equation}
S = - ( \partial F / \partial T)_{V,N} = N [ln (V / N \lambda^3) + 5/2].
\label{entropy1}
\end{equation}

The important point to note is that, since $V \sim \lambda^3$, the entropy
$S$ in Eq. (\ref{entropy1}) becomes nonsensically negative unless $ N \sim
1$ which is equally nonsensical because $N$ should not be too different from
$(R_H/l_P)^2 \gg 1$.
But the solution \cite{plb} 
is obvious: the $N$ inside the log in Eq. (\ref{entropy1}) somehow
must be absent.  Then $ S \sim N
\sim (R_H/l_P)^2$ without $N$ being small (of order 1) and S is non-negative
as physically required.  That is the case if the ``particles" are
distinguishable and nonidentical!  For in that case, the Gibbs $1/N!$ factor
is absent from the partition function $Z_N$, and the entropy becomes
$
S = N[ln (V/ \lambda^3) + 3/2].
$
\\

Now the only known consistent statistics in greater than two space
dimensions
without the Gibbs factor (recall that the Fermi statistics and Bose
statistics give similar results as
the conventional Boltzmann statistics at high temperature)
is infinite statistics (sometimes called
``quantum Boltzmann statistics") \cite{DHR,govorkov,greenberg}.  Thus we
have
shown that the ``particles" constituting dark energy obey infinite
statistics,
instead of the familiar Fermi or Bose statistics \cite{plb}.
\footnote{
Using the Matrix theory approach,
Jejjala, Kavic and Minic \cite{minic} have also argued that dark energy 
quanta obey infinite statistics.}\\

To show that the quanta of modified dark matter also obey
this exotic statistics, we \cite{PRD}  first reformulate MoND via an effective 
gravitational dielectric medium, motivated by the analogy \cite{dielectric} 
between Coulomb's law in a dielectric medium and Milgrom's law for MoND.
Ho, Minic and I then find that MONDian force law is recovered if the quanta
of MDM obey infinite statistics.\\

What is infinite statistics?  Succinctly, a Fock realization of infinite
statistics is provided by a $q = 0$ deformation of the 
commutation relations of
the oscillators:
$a_k a^{\dagger}_l - q a^{\dagger}_l a_k = \delta_{kl}$ with $q$ between -1
and 1 (the case $q = \pm 1$ corresponds to bosons or fermions).  States are
built by acting on a vacuum which is annihilated by $a_k$.  Two states
obtained by acting with the $N$ oscillators in different orders are
orthogonal.  It follows that the states may be in any representation
of the permutation group.  The statistical mechanics of particles obeying
infinite statistics can be obtained in a way similar to Boltzmann
statistics, with the crucial difference that the Gibbs
$1/N!$ factor is absent for the former.  Infinite statistics can be
thought of as corresponding to the statistics of identical particles with an
infinite number of internal degrees of freedom, which is
equivalent to the statistics of nonidentical particles since they are
distinguishable by their internal states.\\

It has been shown that a theory of particles
obeying infinite statistics cannot be local \cite{fredenhagen,
greenberg}.  For example, the expression for the number operator,
\begin{equation}
n_i = a_i^{\dagger} a_i + \sum_k a_k^{\dagger} a_i^{\dagger} a_i a_k 
+ \sum_l \sum_k a_l^{\dagger} a_k^{\dagger} a_i^{\dagger} a_i a_k a_l + ...,
\label{number}
\end{equation}
is both nonlocal and nonpolynomial in the field operators,
and so is the Hamiltonian.  The lack of
locality may make it difficult to formulate a relativistic verion of the
theory; but it appears that a non-relativistic theory can be developed.
Lacking locality also means that the familiar spin-statistics relation is no
longer valid for particles obeying infinite statistics; hence
they can have any spin.  Remarkably, the TCP theorem and cluster
decomposition have been shown to hold despite the lack of locality
\cite{greenberg}.\\

According to the holographic principle, the
number of degrees of freedom in a region of space is bounded not by
the volume but by the surrounding surface.  This suggests that the
physical degrees of freedom are not independent but, considered
at the Planck scale, they must be infinitely correlated, with the result
that the spacetime location of an event may lose its invariant significance.
Since the holographic principle is believed to be
an important ingredient in the formulation of quantum gravity,
the lack of locality for theories of infinite statistics may not be a
defect; it can actually be a virtue.  Perhaps it is this lack of locality 
that makes it possible to incorporate gravitational interactions in
the theory.  Quantum gravity and infinite statistics appear to fit together
nicely, and nonlocality seems to be a common feature of both of them
\cite{plb}.\\

\section{Summary and Conclusion}
The dark sector in the concordant model of cosmology $\Lambda$CDM
has two components: dark energy and dark matter.
We have argued that quantum fluctuations
of spacetime give rise to dark energy in the form of an effective 
cosmological constant $\Lambda$ of the correct magnitude as observed
-- a result also expected for the
present and recent cosmic eras in (generalized) unimodular gravity
and causal-set theory.  In a spacetime with positive
$\Lambda$, gravitational thermodynamics arguments then show that 
dark matter (i.e., modified dark matter)
necessarily exists whose mass profile is intimately
related to $\Lambda$ and ordinary matter, with an emergent 
acceleraton parameter related to $\Lambda$ and the Hubble 
parameter $H$, of the magnitude required to explain
flat galactic rotation curves.  Thus the dark sector in our
Universe may indeed have its origin in quantum gravity.\\

Pursuing this line of argument further, we find that the
quanta of the dark sector appear to obey an unfamiliar
statistics, viz, infinite statistics (or quantum
Boltzmann statistics).  This indicates that 
the dark sector is made up of extended quanta.  As a result,
we expect novel particle phenomenology for interactions 
involving dark matter, thereby ``explaining" why so far 
dark matter detection experiments have not yet convincingly
detected dark matter.  The extended nature of the MDM quanta 
may also explain why the mass profile of MDM depends
on such global aspects of spacetime as $\Lambda$ and $H$.\\


MDM has passed observational tests at both the galactic and cluster 
scales.  We can also mention that
preliminary examinations have
demonstrated (see Ref. \cite{HMN})  
that the cosmology with MDM is well described by the usual 
Friedmann's equations.  We anticipate that this fact 
will allow MDM to predict the correct cosmic microwave 
background (CMB) spectrum shapes as well as the alternating peaks.
And as briefly explained in Ref. \cite{ng16a}, the MDM mass 
distribution as found appears to be consistent with the observed
strong gravitational lensing.\\

We conclude by listing a few items on our lengthy to-do list. 
We plan to study 
concrete constraints from gravitational lensing 
and the bullet cluster on MDM. And 
we would like to answer these questions: 
Can we distinguish MDM from CDM? How strongly
coupled is MDM to baryonic matter?  How does MDM self-interact? We will 
also test MDM at 
cosmic scales by studying the acoustic peaks in the CMB and by doing
simulations of structure formation. 
Last but not least, if the quanta of MDM indeed obey infinite statistics
as we found, can quantum gravity
be the origin of particle statistics and can the underlying 
statistics be infinite statistics such that ordinary particles
obeying Bose or Fermi statistics are actually some sort of 
collective degrees of freedom of more fundamental entities 
obeying infinite statistics?  And if so, what 
are the implications for grand unification?

\section*{Acknowledgments}

This talk is partly based on work done in collaboration with (the late) H. 
van Dam, S. Lloyd, M. Arzano, T. Kephart, C. M. Ho, D. Minic, D. Edmonds,
D. Farrah, and T. Takeuchi.  I thank them all.  The work reported  
here was supported in part by the US Department of Energy, the Bahnson
Fund, and the Kenan Professorship Research Fund of UNC-CH.\\

\end{document}